\documentclass[12pt]{article}

\begin{document}

\author{N. Barat and J.C. Kimball \\
Physics Department, University at Albany\\
Albany, NY 12222}
\date{}
\title{Localization and causality for a free particle}
\date{}
\maketitle

\begin{abstract}
Theorems (most notably by Hegerfeldt) prove that an initially localized
particle whose time evolution is determined by a positive Hamiltonian will
violate causality. We argue that this apparent paradox is resolved for a
positive energy free particle described by either the Dirac equation or the
Klein-Gordon equation because such a particle cannot be localized in the
sense required by the theorems.

PACS: 03.65.Pm, 03.65.Ta, 03.67.Ud, 03.67.Hk

Keywords: Localization, Causality, Hegerfeldt's theorem
\end{abstract}

\section{Introduction}

According to Hegerfeldt's theorem \cite{Heger}, ``Positivity of the
Hamiltonian alone [means] that particles, if initially localized in a finite
region, immediately develop infinite tails.'' Does this mean signals
propagate faster than light \cite{Larpov}? Is causality in jeopardy of being
overthrown, or is this theorem further evidence of cracks in the foundations
of quantum mechanics? In our view, the Hegerfeldt theorem does \textit{not}
imply a failure of causality or relativity. Instead, it is a comment on the
nature of quantum mechanics (but not a fatal flaw in quantum mechanics).
Hegerfeldt's theorem is an ``if...then'' statement, and since the ``then''
part (immediate infinite tails) is nonsensical physics, we conclude that the
``if'' part of the theorem (localized wave functions) should not be
realizable for a sound quantum theory. Thus Hegerfeldt's theorem really
means that a logically consistent single-particle quantum theory should not
allow localization. To back up this claim, we offer analyses of the
free-particle Dirac and Klein-Gordon equations to show that for positive
energy solutions, the traditional definitions of particle density (or energy
density for the Klein-Gordon equation) do not allow the localization
required by the Hegerfeldt theorem.

We will focus our attention on the strongest version of Hegerfeldt's
theorem, presented in Ref.\cite{Heger2}, and referred to as H1. The H1
theorem obtains superluminal speeds if the probability of finding a particle
outside a bounded region decays with distance at a sufficiently large
exponential rate. The results derived here apply to H1 as well as to weaker
versions of the theorem (where more stringent localization is needed to
prove superluminal speed) \cite{Heger3}, \cite{Heger4}. Our results are not
directly related to localization theorems of field theory (see e.g. J.M.
Knight\cite{Knight}).

The building blocks of Hegerfeldt's theorem are a positive Hamiltonian, $H$,
and a localization operator, $N(V)$, whose expectation value corresponds to
the probability of finding a particle (or a particle's energy) inside a
volume $V$. To investigate applicability of the theorem, it is a natural to
consider a Hamiltonian and localization operator based on the most familiar
relativistic wave equations of quantum mechanics: the Dirac and Klein-Gordon
equations. We will consider both these cases.

For a free particle, one can construct a positive Dirac Hamiltonian by
restricting the Hilbert space of wave functions to positive energy
solutions. The simplest definition of $N(V)$ is then obtained by projecting
the standard Dirac probability density onto this positive-energy Hilbert
space.

For the Klein-Gordon equation, the positive energy Hamiltonian is obtained
by taking the square root of the Klein-Gordon equation. The absence of a
positive Klein-Gordon probability density forces us to consider an
alternative interpretation of the Hegerfeldt theorem. If one associates $%
\left\langle N(V)\right\rangle $ with the probability of finding the
particle's \textit{energy} in the region $V$, then $N(V)$ is positive,
normalizable and well-defined. Furthermore, the interpretation of $%
\left\langle N(V)\right\rangle $ as the probability for finding the
particle's energy in the volume $V$ does not change the superluminal
implication of the theorem.

Our basic conclusion is that H1 cannot be applied to the Dirac or
Klein-Gordon equations using the $N(V)$ described above because one cannot
form a sufficiently localized $\left\langle N(V)\right\rangle $ to satisfy
the postulates of the theorem. We view this as a positive result, because it
gives one greater confidence in the internal consistency of these
cornerstones of quantum mechanics. For both the Dirac and Klein-Gordon
equations, our mathematics can be summarized and simplified in terms of the
momentum space wave function, $\psi (\vec{k})$. Localization is inconsistent
with the wave equations because
\begin{eqnarray*}
localization &\Rightarrow &analytic\,\,\psi (\vec{k}) \\
wave\,\,equation &\Rightarrow &nonanalytic\,\,\psi (\vec{k})
\end{eqnarray*}
The ``analytic-nonanalytic'' distinction will be made precise in Sections 2
and 3, where we will discuss the Dirac and Klein-Gordon examples.

One can consider alternative forms for $N(V)$ which allow the localization
needed to apply Hegerfeldt's theorem. One famous alternative is based on the
Newton-Wigner\cite{Newton} position operator, which \textit{does} satisfy
the postulate of H1 (as well as weaker versions of the theorem). In our
opinion, the fact that superluminal velocities would occur if the
Newton-Wigner approach were valid is another argument against Newton-Wigner
localization. The Newton-Wigner approach also encounters difficulty with
Lorentz invariance, and particle conservation. (See Kalnay\cite{Kalnay} and
Rosenstein and Usher \cite{Rosen} for additional discussion).

We need to be precise about the meaning of ``localization.'' In H1
superluminal speed was proved when a particle was exponentially localized so
that the probability of finding the particle outside a sphere of radius $R$
was bounded by $\bar{A}^2\exp (-2\gamma R),$ with $\bar{A}<\infty $ and $%
\gamma >m$. We use units where $\hbar =c=1$, so the particle's mass $m$
corresponds to the inverse of its Compton wave length $\lambda =\hbar /(mc)$%
. We take this bound as our definition of localization and apply it in
Sections 2 and 3. The final section includes some comments and a brief
discussion of other work. In particular, we acknowledge the importance of
work by Thaller \cite{Thaller} in our derivations.

\section{Dirac Equation}

Hegerfeldt's theorem applies to systems with a positive Hamiltonian, but the
Dirac equation has both positive and negative eigenvalues. However, for free
particles, the Hilbert space $\mathcal{H}$ of solutions to the Dirac
equation is the sum of positive energy and negative energy Hilbert spaces.
\begin{equation}
\mathcal{H}=h_{+}+h_{-}
\end{equation}
The Dirac free-particle Hamiltonian is well-defined and positive on $h_{+}$.

The Dirac equation allows the construction of a particle current 4-vector,
whose time-like component is a probability density
\begin{equation}
\rho (\vec{r},t)=\sum_{i=1}^4\psi _i^{*}(\vec{r},t)\psi _i(\vec{r},t)
\label{Rho2}
\end{equation}
where the sum is over spinor components. (In the notation of QED, this is
written as $\rho =\bar{\psi}\gamma _0\psi $.) Associated with this
probability density is an operator $N(V)$ which gives the probability that a
particle is within a volume $V$.
\begin{equation}
N(V)=I_4P(V)
\end{equation}
where $P(V)$ is the projection onto the volume $V$%
\begin{equation}
P(V)=\left\{
\begin{array}{lll}
1 &  & \vec{r}\in V \\
0 &  & \vec{r}\notin V
\end{array}
\right.  \label{P(V)}
\end{equation}
and $I_4$ is the unit $4\times 4$ matrix which operates on the $4-$component
spinors. In the full Hilbert space of 4-component solutions, particles can
be localized. For example, one can pick wave functions which vanish outside
a finite region of space. But these localized states do not lie in the
positive energy space $h_{+}$, and projecting them into $h_{+}$ yields
non-localized wave functions. It is this delocalization which makes the
application of Hegerfeldt's theorem impossible.

The natural generalization of $N(V)$ to $h_{+}$ is
\begin{equation}
N(V)_{+}=P_{+}\,\,N(V)\,\,P_{+}
\end{equation}
where $P_{+}$ is the projection from $\mathcal{H}$ onto $h_{+}$. Then for
any wave function $\psi \in h_{+}$, $P_{+}\psi =\psi $ and
\begin{equation}
\left\langle \psi ,N(V)_{+}\psi \right\rangle =\left\langle \psi ,N(V)\psi
\right\rangle
\end{equation}
gives the probability that a positive energy particle is within the volume $%
V $. Thus when working with wave functions in $h_{+}$, one can use the
definition of Eq. (\ref{Rho2}) and omit the subscript on $N(V)$.

To apply Hegerfeldt's theorem as described in H1, one must first exhibit a
normalized wave function in $h_{+}$ which is localized so that \cite{Heger2}
\begin{equation}
\left\langle \psi ,N(B_r)\psi \right\rangle >1-\bar{A}^2\exp (-2\gamma r)
\end{equation}
where $B_r$ is a sphere of radius $r$ and $\gamma >m$. This inequality can
be satisfied only if there is a wave function in $h_{+}$, each of whose
components satisfies the condition
\begin{equation}
\left| \psi _i(\vec{r},0)\right| <A\exp \left( -\gamma r\right)
\label{Bound}
\end{equation}
for all $\vec{r}$ and a finite $A$. This bound on $\psi _i(\vec{r},0)$
implies a range of analyticity of the Fourier transform of the wave function
components given by
\begin{equation}
\psi _i(\vec{k})=\frac 1{\left( 2\pi \right) ^{3/2}}\int \exp (-i\vec{k}%
\cdot \vec{r})\psi _i(\vec{r},0)\,\,d^3r
\end{equation}
Localization means each $\psi _i(\vec{k})$ is an analytic function of the
vector components of $\vec{k}$ in the complex-plane strip characterized by
\begin{equation}
\left| \mbox{Im}(\vec{k})\right| \leq m  \label{Smooth}
\end{equation}
Rather than presenting a proof of this analytic property, we illustrate the
structure of $\psi _i(\vec{k})$ with a simple example. Assume a $\psi _i(%
\vec{r},0)$ is minimally localized, so that it takes on the value of its
upper bound in Eq. (\ref{Bound}). Then the Fourier integral (obtained most
simply in spherical coordinates where $\,d^3r=2\pi r^2d(\cos \theta )dr$ and
$\vec{k}\cdot \vec{r}=kr\cos \theta $) gives
\begin{equation}
\psi _i(\vec{k})=\frac A{\left( 2\pi \right) ^{3/2}}\frac{8\pi \gamma }{%
\left( k^2+\gamma ^2\right) ^2}
\end{equation}
which is analytic for $\left| \mbox{Im}(\vec{k})\right| <\gamma $, and since
$\gamma >m$ we obtain the condition of Eq. (\ref{Smooth}). If the wave
function is more strongly localized, so the magnitude of $\psi _i(\vec{r},0)$
is decreased, the range of analyticity of $\psi _i(\vec{k})$ can only
increase. A formal analyticity proof is based on Theorem IX.13 of Reed and
Simon \cite{Reed}.

The Dirac equations yields the opposite conclusion on the analyticity of
wave functions. The requirement of positive frequency means the four
components of the wave function are not all independent. For
spin-polarization $\vec{\sigma}$, the components of the momentum space wave
function must satisfy the condition \cite{Thaller}
\begin{equation}
\left(
\begin{array}{l}
\psi _3(\vec{k}) \\
\psi _4(\vec{k})
\end{array}
\right) =\frac{\vec{k}\cdot \vec{\sigma}}{\sqrt{m^2+k^2}+m}\left(
\begin{array}{l}
\psi _1(\vec{k}) \\
\psi _2(\vec{k})
\end{array}
\right)
\end{equation}
The branch cut in $\sqrt{m^2+k^2}$ at $k=im$ means all four components $\psi
_i(\vec{k})$ cannot be analytic when $\vec{k}$ is imaginary with magnitude $%
m $, and this is inconsistent with Eq. (\ref{Smooth}).

This analysis of the wave function's analytic structure shows that
Hegerfeldt's theorem does not apply to positive energy free particles
described by the Dirac equation (with the traditional probability density)
because these particles cannot be described by localized wave functions. The
exponential tail required of positive energy solutions to the Dirac equation
decays too slowly to allow application of the theorem of H1.

\section{Klein-Gordon Equation}

The interpretation of a single spin-zero particle described by the
Klein-Gordon equation presents different problems. A positive probability
density cannot be defined (even for a free particle). However, one can
consider a modification of Hegerfeldt's result which uses the particle's
energy density instead of its probability density as the basis for
constructing an $N(V)$. Since relativity also limits the propagation speed
of energy to be less than the speed of light, the potential contradiction of
quantum mechanics and relativity is still an issue.

The Klein-Gordon equation is
\begin{equation}
-\frac{\partial ^2\psi }{\partial t^2}=-\nabla ^2\psi +m^2\psi  \label{KG}
\end{equation}
The positivity assumption means the only allowed solutions are linear
combinations of positive frequency plane wave states of the form
\begin{equation}
\psi _k(\vec{r},t)=\exp \left( i\vec{k}\cdot \vec{r}-i\omega (k)t\right)
\end{equation}
with
\begin{equation}
\omega (k)=+\sqrt{k^2+m^2}  \label{Omega}
\end{equation}
Thus the Klein-Gordon equation restricted to positive frequencies can be
written as
\begin{equation}
i\frac{\partial \psi }{\partial t}=\sqrt{-\nabla ^2+m^2}\,\,\psi
\label{KGpos}
\end{equation}
where the meaning of the positive energy operator on the right is obtained
from the Fourier transformed expression for $\psi $.

As is well know from the history of the Klein-Gordon equation, one cannot
easily justify $\psi ^{*}\psi $ as a probability density. (However, it is
related to the Newton-Wigner\cite{Newton} approach to localization.) The
standard Klein-Gordon density is the ``charge density''
\begin{equation}
\rho _c=\frac i2\left( \psi ^{*}\frac{\partial \psi }{\partial t}-\psi \frac{%
\partial \psi ^{*}}{\partial t}\right)  \label{rhoc}
\end{equation}
However, $\rho _c$ is not always positive, even for the positive frequency
free-particle solutions to the free-particle Klein-Gordon equation.

Because we cannot identify a probability density with acceptable physical
properties, we propose that the Klein-Gordon energy density is the
appropriate way to characterize particle localization. This energy density
is proportional to
\begin{equation}
T(\vec{r},t)=\left| \vec{\nabla}\psi \right| ^2+\left| \frac{\partial \psi }{%
\partial t}\right| ^2+m^2\left| \psi \right| ^2  \label{EnDen}
\end{equation}
We can normalize $\psi $ so that the integral of $T(\vec{r},t=0)$ over all
space is unity (in a fixed coordinate system). Furthermore application of
the Klein-Gordon equation (for a free particle with positive frequencies)
shows that this integral cannot vary in time, so a convenient normalization
gives
\begin{equation}
\int T(\vec{r},t)\,\,d^3r=1
\end{equation}
for any $t$. With this normalization, we define an operator $N(V)$ with $%
0\leq N(V)\leq 1$ which can be interpreted as the probability that the
particle's energy is confined within the volume $V$, because
\begin{equation}
\left\langle \psi ,N(V)\psi \right\rangle =\int_VT(\vec{r},t)\,\,d^3r
\label{N-T}
\end{equation}
Since the momentum operator $-i\vec{\nabla}$ and the operator $\sqrt{-\nabla
^2+m}$ of Eq. (\ref{KGpos}) are Hermitian, the $N(V)$ of Eq. (\ref{N-T}) for
positive energy states is
\begin{equation}
N(V)=\sqrt{-\nabla ^2+m^2}P(V)\sqrt{-\nabla ^2+m^2}-\vec{\nabla}\cdot \left(
P(V)\vec{\nabla}\right) +m^2P(V)
\end{equation}
where $P(V)$ is the projection onto the volume $V$, as defined in Eq. (\ref
{P(V)}).

If $\left\langle \psi ,N(V)\psi \right\rangle $ could be exponentially
localized, Hegerfeldt's theorem would mean a Klein-Gordon particle's energy
density could propagate faster than light. We show that this is impossible
following essentially the same reasoning as was used for the Dirac equation.
Using Eqs. (\ref{EnDen}, \ref{N-T}) localization would mean both $\psi (\vec{%
r},0)$ and $\partial \psi (\vec{r},0)/\partial t$ should be constrained by
the exponential bound of Eq. (\ref{Bound}). However, as with the Dirac
equation, the bounds impose an analytic structure on the Fourier transform
of the wave functions. Using standard relativistic notation, we write
\begin{equation}
\psi (\vec{r},t)=\left( 2\pi \right) ^{-3/2}\int \left( \frac{\hat{\psi}(%
\vec{k})}{\omega (\vec{k})}\right) \exp \left( i\vec{k}\cdot \vec{r}-i\omega
(k)t\right) d^3k  \label{FT}
\end{equation}
where $\hat{\psi}(\vec{k})$ is the ``momentum-space wave function.'' The
inverse Fourier transform of Eq. (\ref{FT}) gives
\begin{equation}
\frac{\hat{\psi}(\vec{k})}{\omega (\vec{k})}=\left( 2\pi \right) ^{-3/2}\int
\psi (\vec{r},0)\exp \left( -i\vec{k}\cdot \vec{r}\right) d^3r  \label{FT2}
\end{equation}
Differentiating Eq. (\ref{FT}) with respect to time and then taking the
inverse Fourier transform gives
\begin{equation}
\hat{\psi}(\vec{k})=\left( 2\pi \right) ^{-3/2}\int i\frac \partial
{\partial t}\psi (\vec{r},0)\exp \left( -i\vec{k}\cdot \vec{r}\right) d^3r
\label{FT3}
\end{equation}
If both $\psi (\vec{r},0)$ and $\partial \psi (\vec{r},0)/\partial t$
satisfy the localization condition of Eq.(\ref{Bound}), then both $\hat{\psi}%
(\vec{k})/\omega (k)$ and $\hat{\psi}(\vec{k})$ must be analytic functions
of the components of $\vec{k}$ in the strip of the complex plane described
by Eq. (\ref{Smooth}). This is not consistent with the branch cuts in $%
\omega (k)$ at $k=\pm im$ which are displayed in Eq. (\ref{Omega}).

The Klein-Gordon wave equation combined with our definition of $N(V)$ gives
the momentum space wave function an analytic structure which prevents the
localization required by H1. Thus Hegerfeldt's theorem cannot prove
superluminal energy propagation for free positive energy particles described
by the Klein-Gordon equation.

\section{Comments}

An alternative way around the paradox suggested by Hegerfeldt's theorem is
based on physical arguments. Berestetskii, Lifshitz and Pitaevskii \cite{LL}%
, \cite{Landau} point out that single-particle quantum mechanics becomes
inadequate (e.g. pair creation) whenever an experiment is done to show that
a particle is localized in a region smaller than its Compton wave length.
This idea was pursued further by Kaloyerou \cite{Kalo}. Yndurain \cite{Ynd}
and other have used the problems associated with localization as one of the
motivations for abandoning quantum mechanics for field theory. While we do
not question the arguments for field theory, we do feel that the
localization problem continues to deserve attention. Even if ordinary
relativistic quantum mechanics is not a fundamental description of nature,
the applicability and limitations of single-particle theories are of
practical importance.

An early form of Hegerfeldt's theorem has been known for many years, and
many have considered its consequences \cite{Larpov}, \cite{Bracken}, \cite
{Wig1}, \cite{Wig2}. We are not the first to suggest that particles cannot
be localized. Perez and Wilde \cite{Perez} suggested that particles could be
only ``essentially localized'' \cite{Steinmann}, \cite{Haag}. Thaller \cite
{Thaller} noticed that a Dirac particle could not be so strictly localized
that its wave function vanished outside a finite region. Our results are an
extension of Thaller's.

The non-localization of Dirac electrons may appear to contradict the result
of Bracken and Melloy, who in the paper ``Localizing the relativistic
electron'' obtain a sequence of states whose position uncertainty can be
made arbitrarily small \cite{Bracken}. The results are not inconsistent.
Since the Bracken-Melloy sequence is not a Cauchy sequence, the localized
pointwise limit function cannot be treated as a solution to the Dirac
equation.

The problem with the positivity of the Klein-Gordon charge density $\rho _c$
has been frequently discussed when an external potential is added to the
Klein-Gordon equation. Sometimes a negative $\rho _c$ is (dubiously)
attributed to strong-field effects (see for example Ref. \cite{Greiner}).

Our suggestion that the energy density is an appropriate alternative to the
probability density is not new. This situation is commonly acknowledged for
the photon. For example, Akhiezer and Berestetskii \cite{Akhiezer} comment
that ``.. the localization of a photon in a region smaller in order of
magnitude than a wavelength has no meaning, and the concept of probability
density for the localization of a photon does not exist. ... In practice, it
is often sufficient (instead of the equation of continuity for the
probability density) to utilize the equation of continuity for the energy
density.'' We claim that these same comments should apply to a Klein-Gordon
particle, except the size of the position uncertainty is the particle's
Compton wave length rather than the photon's classical wave length. We admit
to a prejudice here. Some argue (e.g. Ref. \cite{Hawton}) that a photon does
have a position operator. If such an operator exists, it is certainly not
simple.

We wish to thank K. Lee, D. Liguori and M. Sadiq for reading and improving
our manuscript.

\end{document}